\pdfoutput=1
\documentclass[12pt,titlepage]{article}

\font\grande=cmr9.5 scaled \magstep4
\font\medio=cmr9.5 scaled \magstep2
\outer\def\beginsection#1\par{\medbreak\bigskip
      \message{#1}\leftline{\bf#1}\nobreak\medskip
\vskip-\parskip
      \noindent}
\usepackage{graphicx} 
\begin{document}
\bibliographystyle {unsrt}

\titlepage

\begin{flushright}
\end{flushright}

\vspace{1cm}
\begin{center}
{\grande Quantum coherence of cosmological perturbations}\\
\vspace{1cm}
 Massimo Giovannini 
 \footnote{Electronic address: massimo.giovannini@cern.ch} \\
\vspace{1cm}
{{\sl Department of Physics, 
Theory Division, CERN, 1211 Geneva 23, Switzerland }}\\
\vspace{0.5cm}
{{\sl INFN, Section of Milan-Bicocca, 20126 Milan, Italy}}
\vspace*{1cm}
\end{center}

\vskip 0.3cm
\centerline{\medio  Abstract}
\vskip 0.1cm
The degrees of quantum coherence of cosmological 
perturbations of different spins are computed in the large-scale limit and compared with the 
standard results holding for a single mode of the electromagnetic 
field in an optical cavity. The degree second-order coherence of curvature 
inhomogeneities (and, more generally, of the scalar modes of the geometry) reproduces faithfully the optical 
limit. For the vector and tensor fluctuations the numerical values of the normalized 
degrees of second-order coherence in the zero-time delay limit are always larger than unity 
(which is the Poisson benchmark value)  but differ from the corresponding expressions
 obtainable in the framework of the single-mode approximation. General lessons are drawn on 
 the quantum coherence of large-scale cosmological fluctuations.
\noindent

\vspace{5mm}
\vfill
\newpage
In quantum optics \cite{QO} the degrees of second-order coherence
are determined by correlating the intensities of the radiation emitted by a source.
This observation is at the heart of the Hanbury Brown-Twiss interferometry which 
is customarily used, with complementary purposes, both in quantum optics \cite{HBT1} and 
in high-energy physics \cite{HBT2}. The logic of interfering intensities (as opposed to amplitudes)
is also applicable to large-scale cosmological perturbations \cite{mg1} where the
second-order interference effects can be used to establish, in a model independent 
manner, the classical or quantum origin of curvature fluctuations which have been 
directly probed by Cosmic Microwave Background experiments \cite{CMB1,CMB2}.

The adjective {\em coherent} is often employed with different meanings: 
a signal is sometimes said to be coherent as opposed to noisy.  
The word coherent may also indicate the correlation between two or more functions although 
the functions themselves may have some random properties. The theory of quantum coherence \cite{QO}, 
originally conceived and developed by various authors including Glauber 
and Sudarshan \cite{COH}, gives an unambiguous answer by relating the coherence properties of a state (or of a source) 
to the minimization of the indetermination relations\footnote{A coherent state (characterized by a Poissonian statistics) 
 is, by definition, coherent to all orders and the corresponding normalized degrees 
 of quantum coherence always coincide with unity.}. The purpose of this analysis is to investigate, 
 in a unified perspective, the quantum coherence of the cosmological perturbations of different spins 
 and relate the obtained results to the conventional quantum optical 
 approach based on the single-mode approximation.

The single-mode approximation is widely used to interpret a variety of observations 
both in Mach-Zender and Hanbury Brown-Twiss interferometry \cite{QO}. 
Since many experiments use plane parallel light beams whose transverse intensity profiles are not important for the measured quantities, 
it is often sufficient in interpreting the data to consider the light beams as exciting a single mode of the field. In this 
sense the viewpoint of quantum optics is {\em exclusive} insofar as  the degrees of first- and second-order coherence are exclusively defined 
for a single mode of the field  and they are given by\cite{QO}:
\begin{eqnarray}
\overline{g}^{(1)}(\tau_{1}, \tau_{2}) &=& \frac{\langle \hat{a}^{\dagger}(\tau_{1}) \, \hat{a}(\tau_{2})\rangle}{\sqrt{\langle \hat{a}^{\dagger}(\tau_{1}) \, \hat{a}(\tau_{1})\rangle}
\, \sqrt{\langle \hat{a}^{\dagger}(\tau_{2}) \, \hat{a}(\tau_{2})\rangle}},
\label{g1qm}\\
\overline{g}^{(2)}(\tau_{1}, \tau_{2}) &=& \frac{\langle \hat{a}^{\dagger}(\tau_{1})  \hat{a}^{\dagger}(\tau_{2}) \, \hat{a}(\tau_{2})\, \hat{a}^{\dagger}(\tau_{1})\rangle}{\langle \hat{a}(\tau_{1}) \, \hat{a}(\tau_{1})\rangle \langle \hat{a}^{\dagger}(\tau_{2}) \, \hat{a}(\tau_{2})\rangle}.
\label{g2qm}
\end{eqnarray}
Equations (\ref{g1qm}) and (\ref{g2qm}) define, respectively, the degrees of first and second-order temporal coherence. The expectation 
values of Eqs. (\ref{g1qm})--(\ref{g2qm}) are evaluated on a specific quantum state. A single-mode coherent state is, by definition, an 
eigenstate of the annihilation operator (i.e. $\hat{a} |\alpha \rangle = \alpha  |\alpha \rangle$). Therefore, in the zero time-delay limit 
(i.e. $\tau_{1}- \tau_{2}\to 0$) Eqs. (\ref{g1qm}) and (\ref{g2qm}) imply $\overline{g}^{(1)} = \overline{g}^{(2)} =1$. According to Glauber theory, the property expressed by Eqs. (\ref{g1qm}) and (\ref{g2qm}) holds, for a coherent state,  to any order (i.e. $\overline{g}^{(1)} = \overline{g}^{(2)} =\,.\,.\,.\, = \overline{g}^{(n-1)} = \overline{g}^{(n)} =1$). For a chaotic state with statistical weights provided by the Bose-Einstein distribution\footnote{To evaluate Eqs. (\ref{g1qm})--(\ref{g2qm}) for a chaotic mixture,  the single mode density matrix in the number basis can be written as $\hat{\rho} = \sum_{n} \, p_{n} \, |\,n \,\rangle\, \langle \, n\,|$ where 
the statistical weights are given by $p_{n} = \overline{n}^{n}/(\overline{n} +1)^{n +1}$ and $\overline{n}$ is the mean 
number of quanta (coinciding with the Bose-Einstein occupation number only in the thermal case). },
Eqs. (\ref{g1qm}) and (\ref{g2qm}) demand $\overline{g}^{(1)}=1$ but $\overline{g}^{(2)}=2$. This result is often dubbed by saying that 
chaotic (i.e. white) light is bunched and it exhibits super-Poissonian statistics \cite{QO,COH}. In the case of a single Fock state we have instead $\overline{g}^{(2)} = (1 - 1/n )< 1$ showing that Fock states always lead to sub-Poissonian behaviour \cite{QO}.  While chaotic light is an example of bunched quantum state (i.e. $\overline{g}^{(2)} > 1$ implying more degree of second-order coherence than in the case of a coherent state), 
Fock states are instead antibunched (i.e. $\overline{g}^{(2)} <1$) 
indicating a degree of second-order coherence smaller than in the case of a coherent state. 

In the zero time-delay limit the degree of second-order coherence of Eq. (\ref{g2qm}) has a simple relation with the variance of the probability distribution
associated with a given quantum state.  Defining $\hat{N}= \hat{a}^{\dagger} \hat{a}$, Eq. (\ref{g2qm}) implies 
$ \overline{g}^{(2)} = (D^2 - \overline{n})/\overline{n}^2$ where $\overline{n} = \langle \hat{N}\rangle $ and $D^2 = \langle \hat{N}^2 \rangle - \langle \hat{N} \rangle^2$.  It is sometimes useful to define the so-called Mandel parameter \cite{QO} whose expression is 
${\mathcal Q} = \overline{n} [ \overline{g}^{(2)} -1] = D^2/\overline{n} -1$. The parameter ${\mathcal Q}$ is directly related to the way second-order correlations are parametrized in subatomic physics \cite{HBT2}. 
In the case of a coherent state we have that  $D^2 = \langle \hat{N} \rangle$ while ${\mathcal Q}=0$; this means that the distribution underlying this state is just the Poisson distribution (i.e. the variance coincides with the mean value). 

For the present ends it is important to appreciate that, in the Heisenberg description, the evolution of cosmological perturbations of different spins can be 
parametrized, in a unified perspective, as:
\begin{eqnarray} 
\hat{a}_{\vec{p}\, \alpha} &=& e^{- i \varphi_{p}}\biggl[ \cosh{r_{p}}\, \hat{b}_{\vec{p}\,\alpha} - e^{i \gamma_{p}} \sinh{r_{p}} \,\hat{b}^{\dagger}_{-\vec{p}\,\alpha}\biggr],
\nonumber\\
\hat{a}_{-\vec{p}\, \alpha}^{\dagger} &=&  e^{i \varphi_{p}}\biggl[ \cosh{r_{p}}\, \hat{b}_{-\vec{p}\, \alpha}^{\dagger} - e^{-i \gamma_{p}} \sinh{r_{p}} \, \hat{b}_{\vec{p}\,\alpha}\biggr].
\label{SQ}
\end{eqnarray}
Since $[\hat{a}_{\vec{k}\, \alpha}, \hat{a}_{\vec{p}\, \beta}^{\dagger}] = \delta_{\alpha\beta}
\delta^{(3)}(\vec{k} - \vec{p})$ and $[\hat{b}_{\vec{k}\, \alpha}, \hat{b}_{\vec{p}\, \beta}^{\dagger}] = \delta_{\alpha\beta}
\delta^{(3)}(\vec{k} - \vec{p})$ the transformation connecting the two sets of creation and annihilation operators is unitary. 
 Equation (\ref{SQ}) describes, in one shot, the scalar, vector and tensor case with the proviso that, for scalar fluctuations, 
the polarization index must be dropped; moreover, in the vector and tensor cases, $\alpha$ corresponds to the two (massless) vector or tensor polarizations, as it will be clear in a moment.  The notations of Eq. (\ref{SQ}) are conventional and the only essential point is that the unitary transformation 
connecting the two sets of creation and annihilation operators must be parametrized by two complex numbers 
\begin{equation}
u_{k}(\tau)=  e^{- i \varphi_{k}(\tau)}\cosh{r_{k}(\tau)}, \qquad v_{k}(\tau) = e^{- i (\varphi_{k}(\tau) -\gamma_{k}(\tau))} \sinh{r_{k}(\tau)},
\label{SQ2}
\end{equation}
subjected to the condition $|u_{k}(\tau)|^2 - |v_{k}(\tau)|^2 =1$. In the 
single-mode approximation described by Eqs. (\ref{g1qm})--(\ref{g2qm}), the expression of Eq. (\ref{SQ}) can be written as $\hat{a} = \cosh{r} \,\hat{b} - \sinh{r} \,\hat{b}^{\dagger}$ where the phases, for the sake of simplicity, have been fixed to zero. Taking the limit of zero time-delay and inserting these expressions in Eq. (\ref{g2qm}) we have that: $\overline{g}^{(2)} = 3 + 1/\overline{n}$ with $\overline{n} = \sinh^2{r}$. 
Whenever $\overline{n} \gg 1$ (which corresponds to the physical situation in a cosmological setting) we have that $\overline{g}^{(2)} \to 3$, as implied 
for a squeezed state in the single-mode approximation \cite{QO}.

In the single-mode approximation the degrees of second-order coherence 
range\footnote{Antibunched sources (like for instance Fock states) will be disregarded since they have no analog in 
cosmological applications where the occupation numbers of the field are always large.} 
between $\overline{g}^{(2)} \to 1 $ (in the case of a coherent state) to $\overline{g}^{(2)} \to 3 $ 
(in the case of a squeezed vacuum state). The single-mode thermal state 
is super-Poissonian (i.e. $\overline{g}^{(2)}\to 2$) but less correlated than a squeezed vacuum state. 
In the case of large-scale cosmological perturbations, thanks to the results of the single-mode approximation, 
the normalized degrees of second-order coherence should all be 
close to $3$ (and anyway greatly exceed $1$). Before the direct scrutiny of this plausible expectation,  
we have to remark that the degrees of coherence and the corresponding Glauber correlation 
functions change in the scalar, vector and tensor cases since they are sensitive to the polarizations.
The Glauber correlation in the scalar case can be written as:
\begin{eqnarray}
&& {\mathcal S}^{(n,m)}(x_{1}, \,.\,.\,.\,x_{n}, \, x_{n+1},\, .\,.\,.\,, x_{n +m}) 
\nonumber\\
&&= \mathrm{Tr}\biggl[ \hat{\rho} \, \hat{q}^{(-)}(x_{1})\,.\,.\,.\, \hat{q}^{(-)}(x_{n})
\, \hat{q}^{(+)}(x_{n+1})\,.\,.\,.\, \hat{q}^{(+)}(x_{n+m})\biggr],
\label{corrS}
\end{eqnarray}
where $x_{i} \equiv (\vec{x}_{i}, \, \tau_{i})$ and $\hat{\rho}$ is the density operator representing the (generally mixed) state of the field $\hat{q}$.  The field operator can always be expressed as $\hat{q}(x) = \hat{q}^{(+)}(x) + \hat{q}^{(-)}(x)$, with $\hat{q}^{(+)}(x)= \hat{q}^{(-)\,\dagger}(x)$. By definition we will have that $\hat{q}^{(+)}(x) |\mathrm{vac} \rangle=0$ and also that  $\langle \mathrm{vac} |\, \hat{q}^{(-)}(x) =0$. 
 It is understood that the state $| \mathrm{vac}\rangle$ minimizes the appropriate Hamiltonian which is, in general, different, 
in the scalar, vector and tensor cases. These issues have been addressed in the current literature 
and more specific discussions can be found, for instance, in \cite{mg1,luk,gr0}.
The mode expansion in the scalar case can be written as:
\begin{eqnarray}
\hat{q}(\vec{x},\tau) = \frac{1}{\sqrt{V}} \sum_{\vec{p}} \hat{q}_{\vec{p}}(\tau)\, e^{- i \vec{p}\cdot \vec{x}},\qquad 
\hat{q}_{\vec{p}} = \frac{1}{\sqrt{2 p}} ( \hat{a}_{\vec{p}} + \hat{a}_{-\vec{p}}^{\dagger}),
\label{AA2}
\end{eqnarray}
where $V$ represents a fiducial (normalization) volume. 
As it is clear from Eq. (\ref{AA2}) we also have that $\hat{q}_{\vec{p}}^{\dagger} = \hat{q}_{- \,\vec{p}}$. 
By switching from discrete to continuous modes the creation 
and annihilation operators obey $[\hat{a}_{\vec{k}}, \hat{a}_{\vec{p}}^{\dagger}] = 
\delta^{(3)}(\vec{k} - \vec{p})$ and the sums are replaced by integrals according to 
$\sum_{\vec{k}} \to V \int d^{3} k/(2\pi)^3$. 

We shall always be concerned with the case of a 
conformally flat background geometry $\overline{g}_{\mu\nu} = a^2(\tau) \eta_{\mu\nu}$ where 
$\eta_{\mu\nu}$ is the Minkowski metric (with signature $(+,\, -,\,-,\,-)$) and $a(\tau)$ is the scale factor in the
conformal time parametrization. The field $\hat{q}$ may represent various physical quantities: it could simply be a spectator field 
(minimally or non-minimally coupled to the conformally flat background geometry) but Eq. (\ref{corrS}) applies to the case of the 
curvature perturbations where $\hat{q} = - \hat{{\mathcal R}} z$  and $\hat{{\mathcal R}}$ 
is the field operator corresponding to the perturbations of the spatial curvature on 
comoving orthgonal hypersurfaces \cite{luk}. 
If the source of inhomogeneity is represented by a scalar degree of freedom $\phi$ (not necessarily identified with the inflaton) 
we will have, as usual, that $z= a \phi^{\prime}/{\mathcal H}$ where ${\mathcal H} = a^{\prime}/a$ and the prime denotes 
a derivation with respect to the conformal time coordinate $\tau$. In a complementary perspective, 
when the source of scalar fluctuations is given by a perfect and irrotational fluid, $z$ will have 
a different analytic expression given by $z= a^2 \sqrt{p_{t}+\rho_{t}}/(c_{st} {\mathcal H})$ 
(as originally suggested by Lukash \cite{luk}) where $(p_{t},\, \rho_{t})$ are the total pressure and energy density 
of the fluid and $c_{st}^2 = p_{t}^{\prime}/\rho_{t}^{\prime}$ is the total sound speed. 
In both cases the evolution of $u_{k}(\tau)$ and $v_{k}(\tau)$ appearing 
in Eq. (\ref{SQ2}) can be written as:
\begin{equation}
u_{k}^{\prime} = - i k \,u_{k} - \frac{z^{\prime}}{z} v_{k}^{\ast}, \qquad v_{k}^{\prime} = - i k\, v_{k} - \frac{z^{\prime}}{z} u_{k}^{\ast},
\label{AA3}
\end{equation}
where, as already mentioned, the prime denotes a derivation with respect to $\tau$.

The Glauber correlation function has been originally defined in the vector case \cite{COH} and in the framework
of quantum electrodynamics; the analog of Eq. (\ref{corrS}) in the vector case is therefore given by:
\begin{eqnarray}
&& {\mathcal V}^{(n,m)}_{i_{1}, \,.\,.\,.\,i_{n}, \, i_{n+1},\, .\,.\,.\,, i_{n +m} }(x_{1}, \,.\,.\,.\,x_{n}, \, x_{n+1},\, .\,.\,.\,, x_{n +m}) 
\nonumber\\
&& = \mathrm{Tr}\biggl[ \hat{\rho} \, \hat{{\mathcal A}}_{i_{1}}^{(-)}(x_{1})\,.\,.\,.\, \hat{{\mathcal A}}_{i_{n}}^{(-)}(x_{n})
\, \hat{{\mathcal A}}_{i_{n+1}}^{(+)}(x_{n+1})\,.\,.\,.\,\hat{{\mathcal A}}_{i_{n+m}}^{(+)}(x_{n+m})\biggr],
\label{corrV}
\end{eqnarray}
where $x_{i} \equiv (\vec{x}_{i}, \, \tau_{i})$ and $\hat{\rho}$ is the density operator representing the (generally mixed) 
state of the field\footnote{The field $\hat{{\mathcal A}}_{i}(\vec{x}, \tau)$ can always be expressed as 
 $\hat{{\mathcal A}}_{i}(x) = \hat{{\mathcal A}}_{i}^{(+)}(x) + \hat{{\mathcal A}}_{i}^{(-)}(x)$, 
 with $\hat{{\mathcal A}}_{i}^{(+)}(x)= \hat{{\mathcal A}}_{i}^{(-)\,\dagger}(x)$.
By definition we will have that $\hat{{\mathcal A}}_{i}^{(+)}(x) |\mathrm{vac} \rangle=0$ 
and also that  $\langle \mathrm{vac} |\, \hat{{\mathcal A}}_{i}^{(-)}(x) =0$;  
the state $|\mathrm{vac}\rangle $ denotes the vacuum.} $\hat{{\mathcal A}}_{i}(\vec{x},\tau)$. The difference between Eqs. (\ref{corrS}) and (\ref{corrV}) is represented by the vector 
polarizations denoted, in Eq. (\ref{corrV}), by $(i_{1}, \,.\,.\,.\,i_{n}, \, i_{n+1},\, .\,.\,.\,, i_{n +m})$.
 With the same notations of Eq. (\ref{AA2}) the mode expansion in the vector case can be written as:
\begin{equation}
\hat{{\mathcal A}}_{i}(\vec{x},\tau) = \frac{1}{\sqrt{V}} \sum_{\vec{p}, \, \alpha} e^{(\alpha)}_{i} \, \hat{{\mathcal A}}_{\vec{p},\, \alpha}(\tau)\, e^{- i \vec{p}\cdot \vec{x}},\qquad  \hat{{\mathcal A}}_{\vec{p},\, \alpha} = \frac{1}{\sqrt{2 p}} ( \hat{a}_{\vec{p}\, \alpha} + \hat{a}_{-\vec{p}\, \alpha}^{\dagger}).
\label{AA4}
\end{equation}
where $e^{(\alpha)}_{i}(\hat{k})$ (with $\alpha = 1, \, 2$) are the two polarizations which are mutually orthogonal and orthogonal to $\hat{k}$; as in the case of Eq. (\ref{AA2}) we have that $\hat{{\mathcal A}}_{\vec{p},\alpha}^{\dagger} = \hat{{\mathcal A}}_{- \,\vec{p},\alpha}$.
Since we shall be mainly concerned with massless and divergence-less vectors the sum over the polarizations will be
given by $\sum_{\alpha} e^{(\alpha)}_{i}(\hat{k}) e^{(\alpha)}_{j}(\hat{k}) = P_{ij}(\hat{k})$ where $P_{ij}(\hat{k}) = \delta_{ij} - \hat{k}_{i} \hat{k}_{j}$. In a cosmological context the vector fluctuations may come either from the metric or from gauge fields \cite{mg1}. In the four-dimensional case there are actually two divergence-less vectors
related to the vector fluctuations fluctuations of the metric; they would correspond to $\delta_{\mathrm{v}} g_{0 i} = - a^2 Q_{i}$ and to 
$\delta_{\rm v} g_{i j} = a^2 ( \partial_{i} W_{j} + \partial_{j} W_{i})$ with $\partial_{i} Q^{i} =0 \partial_{i} W^{i} =0$. These two vectors, however,
are not amplified in the concordance paradigm \cite{CMB1,CMB2} and shall not be specifically considered here. Conversely 
the vector fluctuations possibly relevant for our purposes are the ones of gauge fields; in this case $u_{k}(\tau)$ and $v_{k}(\tau)$ 
obey an equation analog to Eq. (\ref{AA3}) but with $z^{\prime}/z\to \chi^{\prime}/\chi$ where $\chi$ denotes the
susceptibility of the Abelian gauge field (see e.g. \cite{mg1}). 

 As in the scalar and vector cases the (divergenceless and traceless) tensor field $\hat{\mu}_{ij}(\vec{x}, \tau)$ can always 
 be expressed as $\hat{\mu}_{ij}(x) = \hat{\mu}_{ij}^{(+)}(x) + \hat{\mu}_{i j}^{(-)}(x)$ 
 and their mode expansion is\footnote{The tensor modes arise naturally 
in the framework of the concordance scenario since their evolution is not Weyl invariant as established long ago by 
Grishchuk \cite{gr0}. In the case of the tensor modes the evolution of $u_{k}(\tau)$ and $v_{k}(\tau)$ is given by 
Eq. (\ref{AA3}) but with $z^{\prime}/z$ replaced by $a^{\prime}/a= {\mathcal H}$.} 
\begin{equation}
\hat{\mu}_{i j}(\vec{x},\tau) = \frac{\sqrt{2 \ell_{P}}}{\sqrt{V}} \sum_{\vec{p}, \, \alpha} e^{(\alpha)}_{i j} \, \hat{\mu}_{\vec{p},\, \alpha}(\tau)\, e^{- i \vec{p}\cdot \vec{x}},\qquad  \hat{\mu}_{\vec{p},\, \alpha} = \frac{1}{\sqrt{2 p}} ( \hat{a}_{\vec{p}\, \alpha} + \hat{a}_{-\vec{p}\, \alpha}^{\dagger}),
\label{AA6}
\end{equation}
where $ \ell_{P} = 8 \pi G$; following the standard practice, in Eq. (\ref{AA6}) and elsewhere we shall adopt units  $16 \pi G =1$. Furthermore, as in the 
scalar and vector cases we shall have that $\hat{\mu}_{\vec{p},\, \alpha}^{\dagger} = \hat{\mu}_{-\vec{p},\, \alpha}$.
The two polarizations of the gravitons in a conformally flat background geometry are:
 \begin{equation}
 e_{ij}^{(\oplus)}(\hat{k}) = (\hat{m}_{i} \hat{m}_{j} - \hat{n}_{i} \hat{n}_{j}), \qquad 
 e_{ij}^{(\otimes)}(\hat{k}) = (\hat{m}_{i} \hat{n}_{j} + \hat{n}_{i} \hat{m}_{j}),
 \label{ST0}
 \end{equation}
 where $\hat{k}_{i} = k_{i}/|\vec{k}|$,  $\hat{m}_{i} = m_{i}/|\vec{m}|$ and $\hat{n} =n_{i}/|\vec{n}|$ 
denote three mutually orthogonal directions. It follows from Eq. (\ref{ST0}) that $e_{ij}^{(\lambda)}\,e_{ij}^{(\lambda')} = 2 \delta_{\lambda\lambda'}$ while the sum over the polarizations gives:
\begin{equation}
A_{i\,j\,m\,n}(\hat{k}) = \sum_{\lambda} e_{ij}^{(\lambda)}(\hat{k}) \, e_{m n}^{(\lambda)}(\hat{k}) = \biggl[P_{m i}(\hat{k}) P_{n j}(\hat{k}) + P_{m j}(\hat{k}) P_{n i}(\hat{k}) - P_{i j}(\hat{k}) P_{m n}(\hat{k}) \biggr];
\label{ST0B} 
\end{equation}
where, as already mentioned, $P_{ij}(\hat{k}) = (\delta_{i j} - \hat{k}_{i} \hat{k}_{j})$.
In the tensor case the Glauber correlation function is finally given by: 
\begin{eqnarray}
&& {\mathcal T}^{(n,m)}_{(i_{1}\,\,j_{1}), \,.\,.\,.\,(i_{n}\,\,j_{n}), \, (i_{n+1}\,\,j_{n+1}),\, .\,.\,.\,, (i_{n +m}\,\,j_{n+m}) }(x_{1}, \,.\,.\,.\,x_{n}, \, x_{n+1},\, .\,.\,.\,, x_{n +m}) 
\nonumber\\
&& = \mathrm{Tr}\biggl[ \hat{\rho} \, \hat{\mu}_{i_{1}\,\,j_{1}}^{(-)}(x_{1})\,.\,.\,.\, \hat{\mu}_{i_{n}\,\,j_{n}}^{(-)}(x_{n})
\, \hat{\mu}_{(i_{n+1}\,\,j_{n+1})}^{(+)}(x_{n+1})\,.\,.\,.\,\hat{\mu}_{(i_{n+m}\,\,j_{n +m})}^{(+)}(x_{n+m})\biggr],
\label{corrT}
\end{eqnarray}
where $x_{i} \equiv (\vec{x}_{i}, \, \tau_{i})$ and $\hat{\rho}$ is the density operator representing the (generally mixed) state of the field $\hat{\mu}_{ij}$. 
In Eq. (\ref{corrT}) the polarization structure is different from the vector case of Eq. (\ref{corrV}): instead of $n+m$ vector indices we 
have $n+m$ tensor indices. 

The general expressions of the Glauber correlators defined in Eqs. (\ref{corrS}), (\ref{corrV}) and (\ref{corrT}) 
contain a wealth of informations. For the present purposes 
the intensity correlators (also relevant in the context of the Hanbury Brown-Twiss interferometry) can be deduced
 from Eqs. (\ref{corrS}), (\ref{corrV}) and (\ref{corrT}) by identifying the corresponding space-time points as follows:
\begin{equation}
x_{1} \equiv x_{n+1},\qquad x_{2} \equiv x_{n+2},\qquad .\,.\,.\, \qquad x_{n} \equiv x_{2 n}.
\label{ident}
\end{equation}
In this case the original Glauber correlator will effectively be a function of $n$ points and and it will describe the correlation 
of $n$ intensities. Using Eq. (\ref{ident}) into Eq. (\ref{corrS}) in the case $n=2$ we have that the correlation 
of the scalar intensities can be written as:
\begin{eqnarray}
{\mathcal S}^{(2)}(x_{1},\, x_{2})  &=& \langle \hat{{\mathcal I}}_{\mathcal S}(\vec{x}_{1}, \tau_{1}) \, \hat{{\mathcal I}}_{\mathcal S}(\vec{x}_{2}, \tau_{2}) \rangle =
\int \frac{d^{3} k_{1}}{2 k_{1} (2\pi)^3} \int \frac{d^{3} k_{2}}{2 k_{2} (2\pi)^3} 
\nonumber\\
&\times& \biggl[|v_{k_{1}}(\tau_{1})|^2 \, |v_{k_{2}}(\tau_{2})|^2 
+ v_{k_{1}}^{*}(\tau_{1}) v_{k_{1}}(\tau_{2}) v_{k_{2}}^{*}(\tau_{2}) v_{k_{2}}(\tau_{1}) e^{- i (\vec{k}_{1}- \vec{k}_{2}) \cdot\vec{r}}
\nonumber\\
&+& v_{k_{1}}^{*}(\tau_1)\,u_{k_{1}}^{*}(\tau_2) \,u_{k_{2}}(\tau_{1})\, v_{k_{2}}(\tau_{2})\,e^{- i (\vec{k}_{1}+ \vec{k}_{2}) \cdot\vec{r}}\biggr],
\label{scalar2I}
\end{eqnarray}
where $\vec{r} = \vec{x}_{1} - \vec{x}_{2}$. 
Similarly, using Eq. (\ref{ident}) into Eq. (\ref{corrV}), the correlation of the intensities in the vector case is given by
\begin{eqnarray}
{\mathcal V}^{(2)}(x_{1},\, x_{2})  &=& \langle \hat{{\mathcal I}}_{\mathcal V}(\vec{x}_{1}, \tau_{1}) \, \hat{{\mathcal I}}_{\mathcal V}(\vec{x}_{2}, \tau_{2}) \rangle =
\int \frac{d^{3} k_{1}}{2 k_{1} (2\pi)^3} \int \frac{d^{3} k_{2}}{2 k_{2} (2\pi)^3} 
\nonumber\\
&\times& \biggl\{ |v_{k_{1}}(\tau_{1})|^2 \, |v_{k_{2}}(\tau_{2})|^2  P_{ii}(\hat{k}_{1}) P_{jj}(\hat{k}_2) +
\nonumber\\
&+& P_{ij}(\hat{k}_{1}) P_{ij}(\hat{k}_2)  \biggl[ v_{k_{1}}^{*}(\tau_{1}) v_{k_{1}}(\tau_{2}) v_{k_{2}}^{*}(\tau_{2}) v_{k_{2}}(\tau_{1})e^{- i (\vec{k}_{1}- \vec{k}_{2}) \cdot\vec{r}}
\nonumber\\
&+& v_{k_{1}}^{*}(\tau_1)\,u_{k_{1}}^{*}(\tau_2) \,u_{k_{2}}(\tau_{1})\, v_{k_{2}}(\tau_{2})\, e^{- i (\vec{k}_{1}+ \vec{k}_{2}) \cdot\vec{r}}\biggr]\biggr\}.
\label{vector2I}
\end{eqnarray}
Finally, in the tensor case, the analog of Eqs. (\ref{scalar2I}) and (\ref{vector2I}) is given by:
\begin{eqnarray}
{\mathcal T}^{(2)}(x_{1},\, x_{2})  &=& \langle \hat{{\mathcal I}}_{\mathcal T}(\vec{x}_{1}, \tau_{1}) \, \hat{{\mathcal I}}_{\mathcal T}(\vec{x}_{2}, \tau_{2}) \rangle =
\int \frac{d^{3} k_{1}}{2 k_{1} (2\pi)^3} \int \frac{d^{3} k_{2}}{2 k_{2} (2\pi)^3} 
\nonumber\\
&\times& \biggl\{ |v_{k_{1}}(\tau_{1})|^2 \, |v_{k_{2}}(\tau_{2})|^2  A_{i\,j\,i\,j}(\hat{k}_{1}) A_{\ell\,m\,\ell\, m}(\hat{k}_2) 
\nonumber\\
&+& A_{i\,j\,\ell\, m}(\hat{k}_{1})  A_{i\,j\,\ell\, m}(\hat{k}_{2}) \biggl[ v_{k_{1}}^{*}(\tau_{1}) v_{k_{1}}(\tau_{2}) v_{k_{2}}^{*}(\tau_{2}) v_{k_{2}}(\tau_{1}) \,e^{- i (\vec{k}_{1}- \vec{k}_{2}) \cdot\vec{r}}
\nonumber\\
&+& v_{k_{1}}^{*}(\tau_1)\,u_{k_{1}}^{*}(\tau_2) \,u_{k_{2}}(\tau_{1})\, v_{k_{2}}(\tau_{2}) e^{- i (\vec{k}_{1}+ \vec{k}_{2}) \cdot\vec{r}}\biggr]\biggr\}.
\label{tensor2I}
\end{eqnarray}
We are now in condition of comparing the degree of second-order coherence 
obtained in the framework of the single-mode approximation (see Eq. (\ref{g2qm})) with the 
results of Eqs. (\ref{scalar2I}), (\ref{vector2I}) and (\ref{tensor2I}) which determine 
the corresponding degrees of second-order coherence:
\begin{eqnarray}
g^{(2)}_{{\mathcal S}}(x_{1}, x_{2}) &=& \frac{{\mathcal S}^{(2)}(x_{1},\, x_{2}) }{{\mathcal S}^{(1)}(x_{1},\, x_{1}) {\mathcal S}^{(1)}(x_{1},\, x_{1})} = 
\frac{\langle \hat{{\mathcal I}}_{\mathcal S}(\vec{x}_{1}, \tau_{1}) \, \hat{{\mathcal I}}_{\mathcal S}(\vec{x}_{2}, \tau_{2})\rangle}{\langle \hat{{\mathcal I}}_{\mathcal S}(\vec{x}_{1}, \tau_{1})\rangle \langle  \hat{{\mathcal I}}_{\mathcal S}(\vec{x}_{2}, \tau_{2}) \rangle},
\label{secondscal}\\
g^{(2)}_{{\mathcal V}}(x_{1}, x_{2}) &=& \frac{{\mathcal V}^{(2)}(x_{1},\, x_{2}) }{{\mathcal V}^{(1)}(x_{1},\, x_{1}) {\mathcal V}^{(1)}(x_{1},\, x_{1})} = 
\frac{\langle \hat{{\mathcal I}}_{\mathcal V}(\vec{x}_{1}, \tau_{1}) \, \hat{{\mathcal I}}_{\mathcal V}(\vec{x}_{2}, \tau_{2})\rangle }{\langle \hat{{\mathcal I}}_{\mathcal V}(\vec{x}_{1}, \tau_{1})\rangle \langle  \hat{{\mathcal I}}_{\mathcal V}(\vec{x}_{2}, \tau_{2}) \rangle},
\label{secondvect}\\
g^{(2)}_{{\mathcal T}}(x_{1}, x_{2}) &=& \frac{{\mathcal T}^{(2)}(x_{1},\, x_{2}) }{{\mathcal T}^{(1)}(x_{1},\, x_{1}) {\mathcal T}^{(1)}(x_{1},\, x_{1})} = 
\frac{\langle\hat{{\mathcal I}}_{\mathcal T}(\vec{x}_{1}, \tau_{1}) \, \hat{{\mathcal I}}_{\mathcal T}(\vec{x}_{2}, \tau_{2}) \rangle}{\langle \hat{{\mathcal I}}_{\mathcal T}(\vec{x}_{1}, \tau_{1})\rangle \langle  \hat{{\mathcal I}}_{\mathcal T}(\vec{x}_{2}, \tau_{2}) \rangle}.
\label{secondtens}
\end{eqnarray}

In the zero time-delay limit  $\tau_{1}- \tau_{2}\to 0$ the scalar degree of quantum coherence becomes 
\begin{eqnarray}
g^{(2)}_{{\mathcal S}}(\vec{r},\tau) &=&  1 + \frac{\int k_{1} d k_{1} |v_{k_{1}}(\tau)|^2 \, j_{0}(k_{1} r) \,\,\int k_{2} d k_{2} |v_{k_{2}}(\tau)|^2\, j_{0}(k_{2} r) }{\int k_{1} \,d k_{1} |v_{k_{1}}(\tau)|^2\,\int k_{2} \,d k_{2} |v_{k_{2}}(\tau)|^2}
\nonumber\\
&+&  \frac{\int k_{1} d k_{1} \,u_{k_{1}}^{*}(\tau) v_{k_1}^{*}(\tau)\, j_{0}(k_{1} r) \,\,\int k_{2} d k_{2} \, u_{k_{2}}(\tau) v_{k_2}(\tau)\,j_{0}(k_{2} r)}{\int k_{1} d k_{1} |v_{k_{1}}(\tau)|^2 \,\,\int k_{2} d k_{2} |v_{k_{2}}(\tau)|^2},
\label{gescal1}
\end{eqnarray}
where $j_{0}(k_{1} r)$ and  $j_{0}(k_{2} r)$ denote the spherical Bessel function of zeroth order.
In the large-scale limit $k_{1} r \ll 1$ and $k_{2} r\ll1$, Eq. (\ref{gescal1}) implies that $g^{(2)}_{{\mathcal S}}(\vec{r},\tau) \to 3$.
The reason for this result stems from the observation that $u_{k_{1}}^{*}(\tau) v_{k_1}^{*}(\tau) u_{k_{2}}(\tau) v_{k_2}(\tau) =  |v_{k_{1}}(\tau)|^2  |v_{k_{2}}(\tau)|^2[ 1 + {\mathcal O}(k_{1}\tau) + {\mathcal O}(k_{2}\tau)]$. This result can be easily obtained in the case of a quasi de Sitter evolution 
where the solution of Eq. (\ref{AA3}) with the correct boundary conditions implies: 
\begin{equation}
v_{k}(\tau) = - \frac{e^{- i k\tau}}{2 k^2 \tau^2}, \qquad u_{k}(\tau) = e^{- i k \tau} \biggl[ 1 - \frac{i}{k \tau} - \frac{1}{2 k^2 \tau^2}\biggr].
\label{int1}
\end{equation}
In the vector and tensor cases the analysis proceeds along the same lines, with the difference that the angular integrals appearing in Eqs.
(\ref{vector2I}) and (\ref{tensor2I}) are more complicated. This is due to the presence of the terms 
\begin{equation}
P_{ij}(\hat{k}_{1})P_{ij}(\hat{k}_{2}) = 1 + (\hat{k}_{1}\cdot\hat{k}_{2})^2, \qquad 
A_{i\,j\,\ell\, m}(\hat{k}_{1})  A_{i\,j\,\ell\, m}(\hat{k}_{2}) = [1 + (\hat{k}_{1}\cdot\hat{k}_{2})^2][ 1 + 3(\hat{k}_{1}\cdot\hat{k}_{2})^2];
\label{int2}
\end{equation}
needless to say that, by definition, $P_{ii}(\hat{k})= 2$ and that $A_{i j i j}= 4$. 
After performing some lengthy (but straightforward) angular integrals, the results of Eq. (\ref{int2}) imply that, 
in the zero-time delay limit  $g_{{\mathcal V}}^{(2)}(x_{1},x_{2}) $ and $g_{{\mathcal T}}^{(2)}(x_{1},x_{2}) $ go, 
respectively, to $5/3$ and to $71/60$.

All in all Eqs. (\ref{secondscal}), (\ref{secondvect}) and (\ref{secondtens}) in the zero time-delay limit (i.e. 
$\tau_{1} \to \tau_{2}$) and for large-scales imply the following results for the degrees of quantum coherence:
\begin{eqnarray}
\lim_{\tau_{1} \to \tau_{2}, \, k r \ll 1} g_{{\mathcal S}}^{(2)}(r,\tau_{1}, \tau_{2}) &=& g_{{\mathcal S}}^{(2)}(r,\tau) \to 3,
\label{sresult}\\
\lim_{\tau_{1} \to \tau_{2}, \, k r \ll 1} g_{{\mathcal V}}^{(2)}(r,\tau_{1}, \tau_{2}) &=& g_{{\mathcal V}}^{(2)}(r,\tau) \to \frac{5}{3},
\label{vresult}\\
\lim_{\tau_{1} \to \tau_{2}, \, k r \ll 1} g_{{\mathcal T}}^{(2)}(r,\tau_{1}, \tau_{2}) &=& g_{{\mathcal T}}^{(2)}(r,\tau) \to \frac{71}{60}.
\label{tresult}
\end{eqnarray}
The quantum optical results (obtained in the single-mode approximation) is only recovered in the scalar case where the degree 
of second-order coherence goes to $3$ exactly as in the case of a (single-mode) squeezed state. When the vector 
polarizations are taken into account the degree of quantum coherence is always super-Poissonian but is given by $5/3$.
Finally, in the tensor case, the degree of quantum coherence is $71/60$ (i.e. slightly above $1$). 
We have therefore that $ g_{{\mathcal S}}^{(2)}(r,\tau) >  g_{{\mathcal V}}^{(2)}(r,\tau) > g_{{\mathcal T}}^{(2)}(r,\tau)$. 
In the vector and the tensor cases  the degree of second-order 
approaches $1$ from above, as implied by the super-Poissonian character of the original quantum state 
of relic photons \cite{mg1} and of relic gravitons \cite{gr0}.

It is well known that first order interference effects between the amplitudes cannot be used to distinguish the nature of different quantum states of the radiation field. 
Young interferometry is not able, by itself, to provide information on the statistical properties of the quantum correlations since various states with diverse 
physical properties (such as laser light and chaotic light) may lead to comparable degrees of first-order coherence \cite{QO}.
The statistical properties of the quantum states can be disambiguated 
by examining the higher degrees of coherence. This program has been specifically suggested 
also in the case of the large-scale curvature perturbations determining the temperature and polarization 
anisotropies of the Cosmic Microwave Background \cite{mg1}. 
For these analyses the new generations of CMB detectors and the Hanbury Brown-Twiss interferometry in the THz region 
could be a plausible framework. In this respect the present findings suggest that the degree of second-order 
coherence for the curvature perturbations is roughly thrice the one of the relic gravitational waves. 
A full account of this discussion is beyond the scopes of this paper.

The normalized degrees of second-order coherence for the cosmological perturbations have been computed with 
the purpose of comparing the obtained results with the standard quantum optical derivations obtained in the single-mode approximation. 
Since the large-scale cosmological perturbations are typically described by squeezed quantum states, it could be naively expected that 
their degree of second-order coherence should always equal $3$ in spite of the spin of the fluctuation. 
While this is true in the scalar case, for the vector and tensor fluctuations the presence of the polarizations reduces the degree 
of second-order coherence in comparison with the scalar case and brings the result much closer to the Poissonian limit. In quantum optics 
the single-mode approximation is fully justified when the experimental set-up effectively involves a single mode of 
the electromagnetic field in a cavity. In the cosmological context this exclusive perspective 
is not typical and the sum over the polarizations cannot be neglected since the quantum state of large-scale 
fluctuations is generally unpolarized. Taken at face value the results presented here suggest that the single-mode approximation cannot 
be used in a cosmological situation. All in all we can instead observe that the effect of the polarizations is 
a progressive reduction of the degree of second-order coherence. This reduction preserves the super-Poissonian 
character of the quantum state so that the Poissonian limit (typical of the coherent state) is never reached.

The author wishes to thank J. Vigen and T. Basaglia of the CERN scientific information service for their kind assistance.

\end{document}